\documentclass[preprintnumbers,twocolumn,groupedaddress,nofootinbib,aps]{revtex4}

\usepackage{amsmath,amssymb}
\usepackage{graphicx}
\usepackage{color}
\usepackage{hyperref}
\usepackage{url}
\usepackage{slashed}
\usepackage{subfigure}
\usepackage{slashed,bbm}
\usepackage{graphics,psfrag,epsfig}
\usepackage{dsfont}
\usepackage{enumerate}
\newcommand{\RNum}[1]{\uppercase\expandafter{\romannumeral #1\relax}}

\newcommand{\be}{\begin{equation}}
\newcommand{\ee}{\end{equation}}
\newcommand{\bea}{\begin{eqnarray}}
\newcommand{\eea}{\end{eqnarray}}

\newcommand{\s}{\sigma}

\newcommand{\la}{\langle}
\newcommand{\ra}{\rangle}

\renewcommand{\vec}[1]{{\boldsymbol #1}}

\newcommand{\im}[1]{{\text{Im}#1}}
\newcommand{\sgn}[1]{{\text{sgn}(#1)}}

\begin{document}

\title{3D non-Fermi liquid behavior  from 1D quantum critical local moments}

\author{Laura Classen,  Igor Zaliznyak, and Alexei M. Tsvelik}
\affiliation{ Condensed Matter Physics and Materials Science Division, Brookhaven National Laboratory, Upton, NY 11973-5000, USA 
}

\begin{abstract}
We study the temperature dependence of the electrical resistivity in a system composed of critical spin chains interacting with three dimensional conduction electrons and driven to criticality via an external magnetic field.
The relevant  experimental system is Yb$_2$Pt$_2$Pb, a metal where itinerant electrons coexist with localized  moments of Yb-ions which can be described in terms of effective $S=1/2$ spins with dominantly one-dimensional exchange interaction. The spin subsystem becomes critical in a relatively weak magnetic field, where it behaves like a Luttinger liquid.
We theoretically examine a  Kondo lattice with different effective space dimensionalities of the two interacting subsystems. We characterize the corresponding non-Fermi liquid behavior due to the spin criticality by calculating the electronic relaxation rate and the  $dc$ resistivity and establish its quasi linear temperature dependence.
\end{abstract}

\maketitle

Coexistence of conduction electrons and localized magnetic moments leads to fascinating physics driven by a competition between screening of the magnetic moments by the conduction electrons, and strong correlations from interaction between the moments. This competition plays itself differently depending on such factors as symmetry, crystal structure, disorder and space dimensionality\cite{brandt1984,lee1986,stewart2001,gegenwart2008,wirth2016}.
Prototypical materials for the study of this interplay are rare earth based metals containing a band of conduction electrons and localized moments of strongly correlated $f$-electrons \cite{stewart1984}.  If the Kondo screening wins, the low energy sector behaves like a Fermi liquid of heavy quasiparticles. It is usually assumed that he opposite case characterized by a strong  interspin interaction leads to magnetic ordering and decoupling of the two subsystems \cite{doniach1977}. Such a scenario, however, is likely to be substantially modified in the case when the spin subsystem is quasi one dimensional and hence cannot order by itself\cite{coleman2010}. This is exactly the case we are interested in.

In our considerations, we have been motivated by experiments on Yb$_2$Pt$_2$Pb, a  metallic compound, where the itinerant electrons originate from Pt and Pb ions and  the local moments are formed by electrons localized on the orbitals of 4f$^{13}$ Yb ions\cite{poettgen}. In the crystal field, these moments are reduced to the  lowest  $j= \pm 7/2$ Kramers doublets\cite{kim2008,ochiai2011}.
There are several features making  Yb$_2$Pt$_2$Pb very special.  
The local moments form a quasi one-dimensional subsystem of effective spins S=1/2 weakly interacting with three dimensional (3D) conduction electrons.  As was found  in \cite{we}, the magnetic excitations are strongly incoherent, gapped (see also \cite{miiller}) and essentially one dimensional  in zero magnetic field. The Ising-like chain is already ordered at $T=0$ and the effect of the small inter-chain coupling (of primarily dipolar origin) is simply to extend this order to finite temperatures ~ 2K \cite{miiller} without noticeably changing the magnitude of the  order parameter \cite{we}. 
The exchange interaction with the conduction electrons appears to be weak, with no experimental indication of Kondo screening \cite{kim2013}, particularly in zero magnetic field when the spectrum is gapped. However, 
when the field overcomes the gap, the magnetic excitations become critical and the 3D magnetic order is strongly suppressed \cite{miiller},\cite{unpublished}. Then their interaction with the conduction electrons becomes important \cite{ref}. 
The strongest critical fluctuations of  one dimensional magnetic chains embedded in a 3D host are centered not at one wave vector, but on an entire plane in 3D reciprocal space. As a consequence, scattering of electrons with critical fluctuations is strong not just at certain  spots, but over a continuum manifold, making  a significant portion of the Fermi surface ``hot". Hence, we expect a regime with non-Fermi-liquid (NFL) behavior, which persists  until the magnetic field exceeds an upper critical threshold, where the magnetization saturates and the excitation spectrum becomes gapped again. Interestingly, the critical behavior is due to local moments that are coupled to, but not hybridized with itinerant electrons.  Furthermore, it appears in the unique situation of a Kondo lattice where the two interacting subsystems  have different effective space dimensionality.

To characterize the emergent NFL behavior, we perturbatively calculate the leading frequency and temperature dependence of the electronic relaxation rate and $dc$ conductivity. They are strongly affected by  the scattering of electrons on  the critical modes of the Yb spin subsystem.
As an advantage of the one-dimensionality, the exact form of the low energy asymptotics of the spin-spin correlation functions  is known so that we can account for the non-trivial dynamics of the local moments in a clear, analytical way. 
We discover a distinct NFL signature in both, conductivity and relaxation rate,  induced by the staggered part of the correlations among local moments. In contrast, the contribution from the uniform component is subleading. In particular, we find that for an extended magnetic field regime, the resistivity scales linearly with temperature.  
This anomalous temperature dependence exists for all current  directions,  parallel or perpendicular to the Yb-chains. However, we find that  the absolute value of the resistivity parallel to the chains is markedly smaller than the one perpendicular to them due to the one dimensional character of the spin excitations. This agrees with the experimental measurements for Yb$_2$Pt$_2$Pb,
which show an anisotropic resistivity linear in $T$ at intermediate magnetic fields\cite{miiller,unpublished}.

{\bf Model} The neutron-scattering experiments in Yb$_2$Pt$_2$Pb can be described by a model of weakly-interacting, anisotropic spin S=1/2 chains \cite{we}
in terms of  exact expressions of the XXZ spin-1/2 chain  Hamiltonian
with moderate Ising anisotropy leading to a small excitation gap $\Delta$.
Here, we will study the situation when this gap closes due to an external magnetic field and the spins enter a Luttinger liquid phase. Due to the dipolar origin of the  interchain coupling its effect in the absence of order becomes negligible \cite{miiller,unpublished}.
The low-energy spin-spin correlation functions take a universal form and do not depend on the exact microscopic model. The only remaining memory is contained  in the correlation function  exponents as shown below. 

Critical modes appear at different wavevectors $Q$, corresponding  to the uniform and staggered components of the correlation functions \cite{chitra1997,giamarchibook,alexeibook,lecturenotes}.
For correlations parallel to the spin $z$-direction, $\la S^z(\omega,q)S^z(0,0)\ra$, they are around $Q=0$ and $Q=\pi(1\pm 2 M)$, where $M\ll1$ is the induced magnetization, $M\propto\sqrt{h-\Delta}$, due to the magnetic field $h=g\mu_b H$. Critical $\la S^+(\omega,q)S^-(0,0)\ra$ modes appear close to $Q=\pm 2\pi M$ and $Q=\pi$. 
The zero-temperature  dynamical susceptibility corresponding to the uniform and staggered correlation functions for the $S^z$ components takes the form
\begin{align}
\chi_{zz}^u(i\omega,q_\|)&=\chi_{zz,0}^{u,0}\frac{q_{||}^2}{\omega^2+v_s^2q_{||}^2}\label{eq:suscep1}\\
\chi_{zz}^{s}(i\omega,q_\|)&=\chi_{zz}^{s,0}\frac{1}{(\omega^2+v_s^2q_{||}^2)^{1-K}} \label{eq:suscep2} 
\end{align}
where $q$ measures the deviation from the critical wavevector $Q$ and the parameters label spin excitation velocity $v_s$ and effective Luttinger liquid parameter $K$. 
The corresponding $S^+S^-$ correlations are given by
\begin{align}
\chi_{+-}^{u}(i\omega,q_\|)&=\chi_{+-}^{u,0}\frac{-\omega^2+v_s^2q^2}{(\omega^2+v_s^2q_{||}^2)^{2-K-1/(4K)}}\label{eq:suscep3}\\
\chi_{+-}^s(i\omega,q_\|)&=\chi_{+-}^{s,0}\frac{1}{(\omega^2+v_s^2q_{||}^2)^{1-1/(4K)}}\label{eq:suscep4}
\end{align}
where $\chi_{+-}/2=\chi_{xx}=\chi_{yy}$ due to spin rotation symmetry around the external magnetic field. 
The Luttinger liquid parameter $K$ varies with magnetic field and depends on the microscopic model. For the XXZ spin-1/2 chain, it takes values between $K=1/4$ and $1$, but is close to $K=1/2$ for an extended regime \cite{okunishi2007}. 
In summary, the spin action becomes
\be
S_s=\sum_{i=x,y,z} \sum_{\alpha=u,s} T\sum_{i\omega}\int_p S^\alpha_i[\chi^{\alpha}_i]^{-1}S^\alpha_i.
\ee
where $\int_p=\int d^3p/(2\pi)^3$  and $T\sum_{i\omega}$ is a summation over Matsubara frequencies.
For the array of noninteracting chains  the spin susceptibility does not depend on the perpendicular momenta, i.e. $\chi_i^\alpha(p_\|,p_\perp)=\chi(p_\|)$.

The electronic part is assumed to be well described by free electrons in a magnetic field, since the material is an excellent metal
\be
S_\psi=T\sum_{i\omega}\int_p (i\omega-\xi_{p\sigma})\psi^\dagger_{p,\sigma}\psi_{p,\sigma}.
\ee
The dispersion relation $\xi_{p,\sigma}=p^2/(2m)-\mu_\sigma$ accounts for the Zeeman splitting modeled by the spin-dependent chemical potential $\mu_\sigma=\mu-\sigma h$ with $\sigma=\pm$. This defines the Fermi vectors $p_{F\s}=\sqrt{2m\mu_\s}$ and $p_{F}=\sqrt{2m\mu}$.
We will label the directions perpendicular and parallel to the local spin chains by $\vec p_\perp$ and $p_\|$.
Both descriptions for electrons and spins are only valid up to a momentum cutoff $\Lambda$.

We model the coupling of the quasi one-dimensional spinons in the Yb-chains to the 3D electrons as 
\be
S_{c}=J T^2\sum_{ik_0,ik_0'} \int_k \int_{k'} V^*_{\vec k\s}V_{\vec k' \s'} \psi^\dagger_{k\s}\vec\s_{\s \s'}\psi_{k'\s'}\vec S( k- k')
\ee
The coupling $J$ is small and only weakly hybridizes the conduction electrons and the spins. 
The matrix elements are given by the overlaps of the plane waves of conduction electrons with the wave functions of the $f$-electrons, which are described by the effective spins $S=1/2$. 
The $f$-electron wave functions are  spherical harmonics with angular  momentum $L=3, J= 7/2$ and $m_L=\pm 3$. Their spacial parts are $V_{\vec k \pm} = Y_{3,\pm3}\left(\vec k/k\right) \sim (k_\|\pm i k_{\perp,1})^3/{k^3}$
with $\vec k_\perp =(k_{\perp,1},k_{\perp,2})^T$. In these notations, the quantization axis of the Yb moments is along $k_{\perp,2}$.

In summary, the action becomes $S=S_s+S_\psi+S_c$. The model is similar to a generalized spin-fermion model, however, in our case spins disperse only in one direction and they are independent degrees of freedom with  non-trivial dynamics inherited from their local character.

\begin{figure}[t]
\centering
 \includegraphics[width=0.43\textwidth]{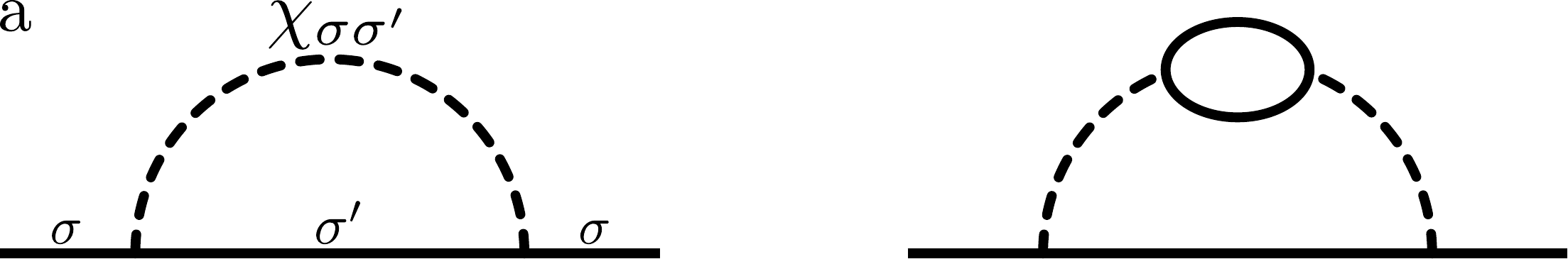} \\ 
\vspace{.5cm}
 \includegraphics[width=0.43\textwidth]{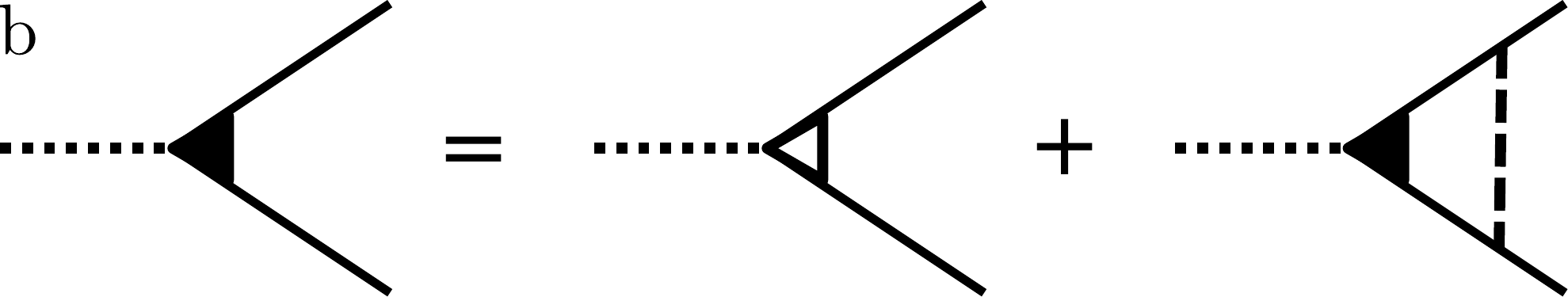}
\caption{(a) Diagrammatic representation of the correction to the electron self-energy without and with renormalized spin propagator. (b) Dyson equation for the current vertex. The empty (black) triangle with the attached dotted line denotes the bare (full) current vertex. Solid (dashed) lines are electron (spin) propagators.}
\label{fig:diagrams}
\end{figure}

{\bf Non-Fermi-liquid regime} The scattering between electrons and spinons leads to remarkable modifications in their correlation functions. In particular, as we will demonstrate, it can destroy the Fermi liquid character of the conduction electrons.
The corresponding information is encoded in the imaginary part of the electron self-energy - the relaxation rate. 
The general form of the one-loop electron self-energy is shown in Fig.~\ref{fig:diagrams} and reads
$ \Sigma_\s(i\omega,\vec k)=
 J^2 \sum_{i,\alpha} T\sum_{i\nu} \int_q |V_{k\s}|^2 |V_{k+q\s'}|^2 \chi_i^\alpha(i\nu,q_\|) G_{\s'}(i\omega+i\nu, \vec k+\vec q)$
 with the bare electron propagator $G_\s^{-1}(i\omega,k)=i\omega-\xi_{\vec k\s}$. In principle, we have to sum over all critical spin modes $\chi_i^\alpha$ in this expression, but we found that the leading contribution comes from the staggered modes  ($\chi_{zz}^s$ and $\chi_{+-}^s$) and we will focus on them in the following. The one-loop self-energy due to the uniform modes is given in the supplemental material (SM). 
 
\begin{figure}[t]
\centering
\includegraphics[width=0.48\textwidth]{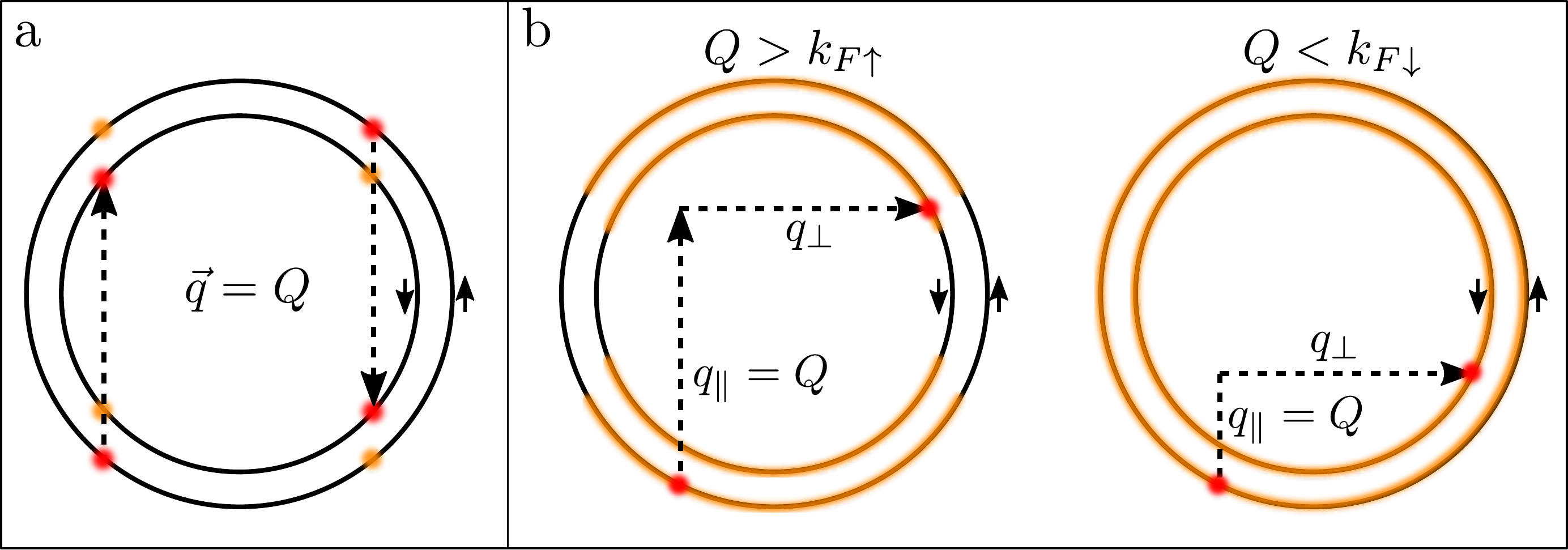}
\caption{(a) If spin and electron correlations are both three dimensional, only a few hot spots or lines appear (colored bright points), where the initial state $k$ and the final state $k+Q$ are both at the FS. (b) If spins are quasi one dimensional and electrons three dimensional a significant part of the FS is hot (colored bright part in the sketch). The reason is that for almost every initial $k$ on the FS, one can find a final $k+q$ that is also on the FS due to the free choice of $q_\perp$. For a spherical FS, it depends on the ratio $Q/k_F$ which part of the FS is hot. If $Q<k_F$ the whole FS is hot.}
\label{fig:hot}
\end{figure}

A special feature of our model is that due to the one dimensional character of the spin excitations these staggered modes affect a significant part of the FS. The dominant contribution to the self-energy on the FS comes from states, where the propagators $G$ and $\chi$ are singular, i.e. $k$ and $k+q$ are close to the FS and $q$ is close to $Q$. Typically for large $ Q$, only small parts of the FS become ``hot'' so that their effect is limited. 
But due to the one-dimensional spin excitations, 
only the parallel component $q_\|\approx Q$ is restricted in our case and we can use the perpendicular component $q_\perp$ to reach a final state on the FS.  This defines an entire  continuum manifold of hot states on the FS. The mechanism is sketched in Fig.~\ref{fig:hot}.

In the following, we will be interested in the frequency dependence of the self-energy. It is thus sufficient to consider the FS-averaged relaxation rate 
\begin{align}
\Gamma_\sigma(\omega)=-\int_p \frac{\delta(\xi_{\vec p \s})}{N_\s} \im{\Sigma}^R_\sigma(\omega,\vec p),
\end{align}
with the retarded self-energy $\Sigma^R(\omega,\vec p)$
 and the density of states at the $\sigma$-FS $N_\s=p_{F\s}m/(2\pi^2)$.

 The resulting FS-averaged relaxation rate due to scattering with staggered spin modes is
\be \label{eq:SE}
\Gamma_\sigma(\omega)\approx N_{\s} C_{\s\s} \omega^{2K} + N_{\bar\s} C_{\s\bar\s} \omega^{\frac{1}{2K}}
\ee
with $\bar\s=-\s$ and the prefactors including the averaged coupling
$C_{\s\s'}=\frac{\chi_{\s\s'}^0\sin(\gamma_{\s\s'}\pi)}{v_s(1-\gamma_{\s\s'})}\widehat{ V^2}$ and 
$\widehat{ V^2}=J^2\int_p \int_q \frac{\delta(\xi_{\s p})\delta(\xi_{\s'p+q+Q_{\s\s'}})}{N_\s N_{\s'}} |V_{\s p}|^2|V_{\s' p+q+Q_{\s\s'}}|^2$.
In this notation, $\chi_{\s\s}=\chi_{zz}$, $\chi_{\s\bar\s}=\chi_{+-}$ and $\gamma_{\s\s}=1-K$, $\gamma_{\s\bar\s}=1-1/(4K)$. Details of the calculation are given in the SM.  
We find that the scaling of the averaged relaxation rate depends on the Luttinger parameter $K$, which varies with the magnetic field. 

Additionally, we can include the electron dynamics into the spin susceptibility and compute the electron self-energy with the renormalized spin propagator (see Fig.~\ref{fig:diagrams}).
 In contrast to the ``conventional'' spin-fermion model, the spins in our heavy-fermion system are independent degrees of freedom, not composite quasiparticles from electron-hole excitations. 
 This means that the corrections to the spin susceptibility become important for energies much smaller than the Fermi energy $\omega< v_s\Lambda \ll E_F$. We are, however, interested in exactly this low-energy regime.
The electron-renormalized spin propagator becomes $\chi^{-1}(\omega,q)=\chi^{-1}_{bare}(\omega,q)+\Pi(\omega,q)$. The real part of the polarization operator $\Pi'$ is approximately constant. The imaginary part depends on the characteristic momentum transfer mediated by the spins. For large momentum transfers it takes the form \cite{abrikosov}
$\Pi(\omega,q)\approx -\Pi_r - i\tau_{\vec q_\perp} \omega,$
where the momentum-dependence of the imaginary part is mainly due to perpendicular momenta as indicated by the subscript because $q_\|\approx Q$.
The leading contribution to the electron scattering rate incuding the polarization bubble in the spin correlation functions is then modified to (see SM)
\begin{align}\label{eq:SEwPi}
 \Gamma_\s(\omega)=\sum_{\s'} N_{\s'} C_{\s\s'} \omega.
\end{align}
We see that, if one takes into account the renormalization of the spin susceptibility by the conduction electrons, the result shows a linear scaling, independent of the value of the Luttinger liquid exponent.

{\bf Conductivity} The temperature dependence of the resistivity is one of the hallmarks of NFL behavior. In a Fermi liquid, it is typically quadratic with temperature. In contrast, in this section, we show  that the resistivity scales linearly with $T$ in the field-induced critical regime
Furthermore, since the scattering is caused by the one-dimensional chains, the coefficient of the linear $T$-term turns out to be different for the directions parallel and perpendicular to the chains. 
For simplicity, we will perform the calculations at zero temperature and assume  that the scaling with temperature is the same as the scaling with frequency (see SM). 

The $dc$ conductivity can be expressed in terms of the Kubo formula as
\begin{align}\label{eq:cond}
\sigma_{\|/\perp}^\sigma\approx&\frac{6e^2n_\s}{c_{\|/\perp}m}\int \frac{d\epsilon}{2\pi}\left(-\frac{dn_F }{d\epsilon}\right) \frac{\gamma_\sigma^{\|/\perp}(\epsilon)}{\Gamma_\s(\epsilon)}
\end{align}
with $c_i=1+\delta_{i,\perp}$ and the number density $n_\s=p_{F\s}^3/(6\pi^2)$. $\gamma_\s$ denotes the scalar part of the current vertex $\vec \Gamma^\s=\vec p \gamma_\s$ and we define its FS-average for the different directions $\gamma_\sigma^{\|/\perp}(\epsilon)=\frac{1}{N_\sigma} \int_p \frac{p_{\|/\perp}^2}{p^2} \delta(\xi_{\sigma\vec p}) \gamma_\sigma(\vec p, \epsilon-i0^+, \epsilon+i0^+)$.
 Their sum, in turn, is the average of the scalar current vertex $\gamma_\sigma(\epsilon)=\gamma^\|_\sigma(\epsilon)+\gamma^\perp_\sigma(\epsilon)=\frac{1}{N_\sigma} \int_p \delta(\xi_{\sigma\vec p}) \gamma_\sigma(\vec p, \epsilon-i0^+, \epsilon+i0^+)$. The approximations used 
 here and in the following are explained in the SM. They only affect prefactors, but lead to the correct scaling behavior \cite{bharadwaj2014,belitz2010}. 

For the correct scaling of the conductivity, vertex corrections are usually important because they weight the scattering events appropriately \cite{Mahan}. The vertex corrections can be determined from the Dyson equation for the current vertex, whose diagrammatic representation is shown in Fig.~\ref{fig:diagrams}.
This leads to
\begin{align}\label{eq:dysoncurr}
\vec p_i^2&\gamma_\sigma(\vec p,ip,ip+i\omega)\notag \\
&=\vec p_i^2+ J^2 \sum_{\sigma'} T\sum_{iq}\int_q |V(p)|^2 |V(q)|^2 \chi_{\sigma\sigma'}(iq-ip,\vec q-\vec p) \notag \\
 &\times G(iq,\vec q)G(iq+i\omega,\vec q) \vec p_i\vec q_i \gamma_{\sigma'}(\vec q,iq,iq+i\omega)
\end{align}
with $i=\|,\perp$.
The resulting scalar current vertex is constant to the leading order:
\begin{align}
\gamma_\sigma&\approx \frac{1}{1-f}+\mathcal{O}(h/E_F),
\end{align}
with $f= 1-Q^2/(2p_{F}^2)+\mathcal{O}(h/E_F)$ and we have neglected terms of order of the magnetic field over the Fermi energy $h/E_F\ll1$. The full expression is given in the SM. 
 As used before, $Q$ is the typical momentum transfer parallel to the chains. To estimate the relative magnitude of $\gamma_\s^\|$ and $\gamma_\s^\perp$, we solve Eq.~\eqref{eq:dysoncurr} for $\gamma_\s^\perp$. In the integral, $\chi$ only depends on $q_\|$ and $p_\|$ and the interaction and the Green's functions are 
 even with respect to $\vec q_\perp\rightarrow -\vec q_\perp$. In contrast, $\vec p_\perp \vec q_\perp$ is odd. Thus the constant 
\begin{align}
\gamma_\s^\perp=\frac{1}{N_\s}\int_p\delta({\xi_{\s\vec p}})\frac{p_\perp^2}{p^2}=\frac{2}{3}
\end{align}
is the solution of  the equation for the vertex $\gamma_\s^\perp$. This result comes from the one-dimensionality of the spin correlations and is independent of the exact form of the interaction $|V|^2$ as long as the interaction is  symmetric under inversion. It further follows that
\begin{align}
\gamma_\s^\|=\gamma_\s-\gamma_\s^\perp\approx \frac{2p_F^2}{Q^2}-\gamma_\s^\perp.
\end{align}
Thus, we find that the direction-resolved scalar current vertices are constant to leading order. Then the scaling of the conductivities is given by the scaling of the relaxation rate. The reason for this is, on the one hand, that vertex corrections for $\gamma_\s^\perp$ are negligible and on the other hand, that $Q$ is finite (for $Q\rightarrow 0$, $f\rightarrow 1$ and $\gamma_\s$ is no longer constant).

In summary, we obtain for the different resistivities
\begin{align}
\rho^{\|,\perp}_\s&=\frac{1}{\sigma_{\|,\perp}^\s}\propto \frac{c_{\|,\perp}}{\gamma_\sigma^{\|,\perp}} \Gamma_\s(T)
\end{align}
This means linear scaling in temperature independent of  $K$ if the polarization $\Pi$ 
 is added to the calculation and $\Gamma_\s(T)\propto T$. For higher energies, when the renormalization by $\Pi$ can be neglected, we found for the leading behavior $\Gamma_\s\propto T^{2K}$ if $1/4<K<1/2$ and $\Gamma_\s\propto T^{\frac{1}{2K}}$ if $1/2<K<1$. In this case, we still obtain approximately linear scaling, because for the Heisenberg-Ising chain $K\approx1/2$ for a significant part of the critical field regime \cite{okunishi2007}. Furthermore we can estimate the ratio of parallel and perpendicular resistivity, which depends on the ratio of $Q$ to $p_F$. This, in turn, is a measure for the ``hotness'' of the FS (Fig.~\ref{fig:hot}). For example, if a dominant part of the FS is hot and $Q\lesssim p_F$, we find
\be
\rho_\|<\rho_\perp/4.
\ee
The relative smallness of the parallel resistivity can be traced back to the negligible corrections of the current vertex perpendicular to the chains, 
which vanish because of the different dimensionality of the conduction electrons and the local moments.

{\bf Conclusion} We have studied a field-induced NFL regime motivated by the heavy-fermion metal Yb$_2$Pt$_2$Pb. 
In this material, there are two weakly coupled subsystems - itinerant electrons from Pb and Pt and localized moments from Yb atoms. The low energy physics 
are determined by the fact that  these subsystems have different effective space dimensionality: magnetic excitations can be described by one-dimensional spin-1/2 chains, while conduction electrons disperse in three dimensions \cite{we}. 
In a sufficiently strong magnetic field, 
the magnetic excitations become critical. 
Here the difference in the dimensionalities comes to the fore because the spin fluctuations being one-dimensional are critical on an entire plane in 3D momentum space. This greatly strengthens the scattering of the conduction electrons. 
We have calculated the resulting electronic relaxation rate and dc conductivity 
 using the exact expressions for the critical spin correlations functions in 1D and found a clear NFL regime for intermediate magnetic fields. In this regime,  where the spins are in the Luttinger liquid phase, the relaxation rate and conductivity scale approximately linearly with frequency or temperature due to 
 scattering on staggered spins fluctuations. 
Furthermore we have given an estimate for the anisotropy of the resistivities parallel and perpendicular to the chains. The parallel resistivity is markedly smaller, because vertex corrections for the perpendicular resistivity vanish. 

In summary, we expect the following behavior for Yb$_2$Pt$_2$Pb in a magnetic field in agreement with 
experiment \cite{unpublished}. At small magnetic fields, the spin excitations are gapped and scattering off them is suppressed at low temperatures. The conductivity can be understood in terms of a conventional  Fermi-liquid. With increasing magnetic field, the gap closes 
and the electron dynamics is altered through scattering with the spins as revealed by the NFL-scaling of the conductivity. Increasing the magnetic field further leads to saturation of the magnetization of the spins. A new gap emerges in their excitation spectrum and we return to the Fermi liquid description.  

Our results characterize a novel form of criticality in a system with coupled local and itinerant degrees of freedom beyond the Doniach paradigm, with different, effective space dimensionality of the subsystems. Besides their general interest with regard to criticality due to localized moments, they could also be of relevance for further compounds of the series R$_2$T$_2$M (R = rare earths or actinides; T = transition metals; M = Cd, In, Sn, and Pb) and other systems in the regime of weak Kondo screening. In general, it would be exciting to identify other materials where the coupling to a critical subsystem with different effective dimensionality induces NFL behavior in the electronic sector.

\begin{acknowledgments}
We acknowledge valuable discussions with Andrey Chubukov, Liusuo Wu, Meigan Aronson and Neil Robinson. L.C. acknowledges support from the Alexander-von-Humboldt foundation.
Work at BNL is supported by the U.S. Department of Energy (DOE), Division of Condensed Matter Physics and Materials Science, under Contract No. DE-SC0012704. 
\end{acknowledgments}

\newpage

\begin{widetext}
\begin{appendix}

\section{Details of the calculations}

\subsection{Spin susceptibilities at finite temperature}\label{app:tempdep}
For completeness we give the spin correlation functions for finite temperature. The susceptibility of commensurate, uniform $S_zS_z$  modes is unchanged compared to Eq.~(2) in the main text.
The remaining susceptibilities are
\begin{align}
\chi_{z}^{ic,R}(\nu,q)&=\chi_{z,0}^c T^{-2 \gamma_{\s\s}}B\Big( \frac{1-\gamma_{\s\s}}{2}-i\frac{\nu+v_s q}{4\pi T},\gamma_{\s\s} \Big) B\Big( \frac{1-\gamma_{\s\s}}{2}-i\frac{\nu-v_s q}{4\pi T},\gamma_{\s\s} \Big)\\
\chi_{+-}^{c,R}(\nu,q)&=\chi_{+-,0}^c T^{-2\gamma_{\s\bar\s}}B\Big( \frac{1-\gamma_{\s\bar\s}}{2}-i\frac{\nu+v_s q}{4\pi T},\gamma_{\s\bar\s} \Big) B\Big( \frac{1-\gamma_{\s\bar\s}}{2}-i\frac{\nu-v_s q}{4\pi T},\gamma_{\s\bar\s} \Big) \\
\chi_{+-}^{ic,R}(\nu,q)&=\chi_{+-,0}^{ic} T^{2-2\beta}\left[B\Big( \frac{3-\beta}{2}-i\frac{\nu+v_s q}{4\pi T},\beta-2 \Big) B\Big( \frac{1-\beta}{2}-i\frac{\nu-v_s q}{4\pi T},\beta \Big) \right. \notag \\
& \left.\hspace{2.5cm}+ B\Big( \frac{1-\beta}{2}-i\frac{\nu+v_s q}{4\pi T},\beta \Big) B\Big( \frac{3-\beta}{2}-i\frac{\nu-v_s q}{4\pi T},\beta-2 \Big)\right]
\end{align}
with $\gamma_{\s\s}=1-K$, $\gamma_{\s\bar \s}=1-{1}/{(4K)}$ and $\beta=2-K-1/(4K)$. The Beta function is given by $B(x,y)=\Gamma(x)\Gamma(y)/\Gamma(x+y)$ with the Gamma function $\Gamma$.
We see that they scale with temperature as with frequency: $T^{-2\gamma_{\s\s'}}$ and $T^{2-2\beta}$ and the remaining dependence is of the form $\nu/T$. Thus, we can obtain the leading temperature scaling of the relaxation rates for vanishing frequency by performing the calculations at zero temperature and replacing the resulting scaling with frequency by the one with temperature.

\subsection {General form of the electron self-energy}\label{app:SE}
We are interested in how the coupling between spinons and electrons affects electronic properties. Therefore we will calculate the electron self-energy $\Sigma_{\sigma}(i\omega,k)$ defined through the propagator $G^{-1}_\sigma(i\omega,k)=i\omega-\xi_{\sigma,\vec k}-\Sigma_\sigma(i,\omega)$. 
The self-energy is given by
\begin{align}
 \Sigma_{\sigma}(i\omega,\vec k)&=J^2 \sum_{i,\sigma'}T \sum_{i\nu} \int \frac{d^3 q}{(2\pi)^3}|V(k)|^2|V(q+k)|^2 \chi_{\sigma \sigma'}^i(\nu,q_\|)G(i\omega+i\nu,\vec k + \vec q) \notag\\
 &=J^2 \sum_{i,\sigma'} T \sum_{i\nu} \int \frac{d^3 q}{(2\pi)^3} \int \frac{dx}{\pi} |V(k)|^2|V(q+k)|^2 \im{\chi_{\sigma \sigma'}^{i,R}(x,q_\|)} \frac{1}{i\nu-x} G(i\omega+i\nu,\vec k + \vec q) \notag\\
 &=J^2 \sum_{i,\sigma'}  \int \frac{d^3 q}{(2\pi)^3} \int \frac{dx}{\pi} |V(k)|^2|V(q+k)|^2\im{\chi_{\sigma \sigma'}^{i,R}(x,q_\|)} \frac{1}{i\omega+x-\xi_{\sigma'}(\vec k
 + \vec q)}\Big( n_B(x) + n_F(\xi_{\sigma'}(\vec k + \vec q)) \Big)
 \end{align}
with $|V(p)|^2=\left( \frac{p_\|^2 + p_{\perp,1}^2}{p^2} \right)^3$. Then the imaginary part of the retarded self-energy becomes
 \begin{align}
 \im \Sigma^R_{\sigma}(\omega,\vec k)&=-J^2 \sum_{i,\sigma'}  \int \frac{d^3 q}{(2\pi)^3} \int \frac{dx}{\pi}|V(k)|^2|V(q+k)|^2 \im{\chi_{\sigma \sigma'}^{i,R}(x,q_\|)} \pi\delta\Big(\omega+x-\xi_{\sigma'}(\vec k + \vec q)\Big)\left( n_B(x) + n_F(\xi_{\sigma'}(\vec k + \vec q)) \right) \notag \\
 &=-J^2 \sum_{i,\sigma'}  \int \frac{d^3 q}{(2\pi)^3}|V(k)|^2|V(q+k)|^2 \im{\chi_{\sigma \sigma'}^{i,R}(\xi_{\sigma'}(\vec k + \vec q)-\omega,q_\|)} \Big(n_B(\xi_{\sigma'}(\vec k + \vec q)-\omega) + n_F(\xi_{\sigma'}(\vec k + \vec q) \Big) \notag \\
  &=-J^2 \sum_{i,\sigma'}  \int \frac{d^3 q}{(2\pi)^3}|V(k)|^2|V(q)|^2 \im{\chi_{\sigma \sigma'}^{i,R}(\xi_{\sigma'}(\vec q)-\omega,q_\|-k_\|)} \Big(n_B(\xi_{\sigma'}(\vec q)-\omega) + n_F(\xi_{\sigma'}(\vec q) \Big).
 \end{align}
The dominant contribution to the self-energy comes from states in the vicinity of the FS. If the momentum transfer $q-p$ is small, this is true for the whole FS. 
In case $q_\|-p_\|\sim Q$ is large, this still concerns a significant part of the FS, because when the final $q_\|\neq k_F'$, we can use the transverse component $q_\perp$ to reach a final state on the FS (see Fig.~2 in the main text).
In the following, we will consider the FS-average of the relaxation rate and use that initial and final states close the the FS give the dominant contribution. We define
\begin{align}
\Gamma_\sigma(\omega)&=-\frac{1}{N_\sigma} \int_p  \delta(\xi_{\sigma\vec p})\im{\Sigma_\sigma^R(\vec p,\omega)} \notag \\
&=J^2 \sum_{i,\sigma'}  \int_p \int_q  \frac{\delta(\xi_{\sigma\vec p})}{N_\sigma} |V(p)|^2|V(q)|^2 \im{\chi_{\sigma \sigma'}^{i,R}(\xi_{\sigma'}(\vec q)-\omega,q_\|-p_\|)} \Big(n_B(\xi_{\sigma'}(\vec q)-\omega) + n_F(\xi_{\sigma'}(\vec q) \Big) \notag \\
&\approx J^2 \sum_{i,\sigma'} {N_{\sigma'}} \int_p \int_q  \frac{\delta(\xi_{\sigma\vec p}) \delta(\xi_{\sigma'\vec q}) }{N_\sigma N_{\sigma'}} |V(p)|^2|V(q)|^2 \int d\xi \im{\chi_{\sigma \sigma'}^{i,R}(\xi-\omega,q_\|-p_\|)} \Big(n_B(\xi-\omega) + n_F(\xi )\Big) \notag \\
&= J^2 \sum_{i,\sigma'} {N_{\sigma'}} \int du \Big(n_B(u) + n_F(u+\omega)\Big) \int_p \int_q  \frac{\delta(\xi_{\sigma\vec p}) \delta(\xi_{\sigma'\vec q})  }{N_\sigma N_{\sigma'}} |V(p)|^2|V(q)|^2  \im{\chi_{\sigma \sigma'}^{i,R}(u,q_\|-p_\|)}  \notag \\
&= J^2 \sum_{i,\sigma'} {N_{\sigma'}} \int du \Big(n_B(u) + n_F(u+\omega)\Big) \int_p \int_q  \frac{\delta(\xi_{\sigma\vec p}) \delta(\xi_{\sigma'\vec q+\vec p+Q})  }{N_\sigma N_{\sigma'}} |V(p)|^2|V(q+p+Q)|^2  \im{\chi_{\sigma \sigma'}^{i,R}(u,q_\|+Q)}  \notag \\
&=:\sum_{i,\sigma'}N_{\sigma'}\int du \Big(n_B(u) + n_F(u+\omega)\Big) \Gamma^i_{\sigma \sigma'}(u)
\end{align}
with $Q$ parallel to the chain-direction and $\int_p=\int d^3p/(2\pi)^3$. The density of states for a spherical FS is $N_\sigma=p_{F\sigma}m/(2\pi^2)$.
Due to simplicity, we will perform the following calculations at zero temperature and then assume that the scaling with frequency is the same as the one with temperature. For zero temperature $n_B(u)+n_F(u+\omega)\rightarrow (\sgn{u}-\sgn{u+\omega})/2$.

\subsection{Relaxation rate due to staggered modes and comparison with numerical evaluation}
\label{app:staggered}
\begin{figure}[t]
\centering
 \includegraphics[width=0.4\textwidth]{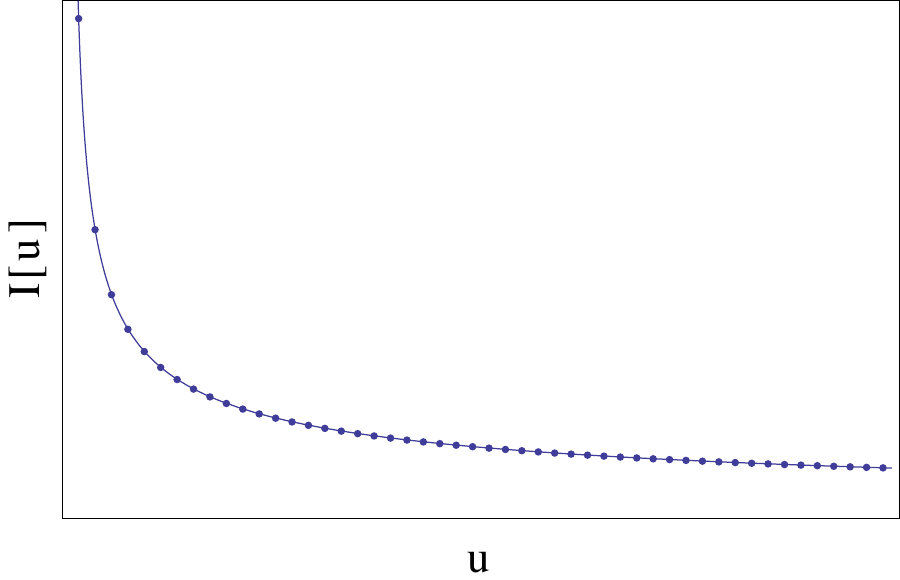} \quad \quad
  \includegraphics[width=0.4\textwidth]{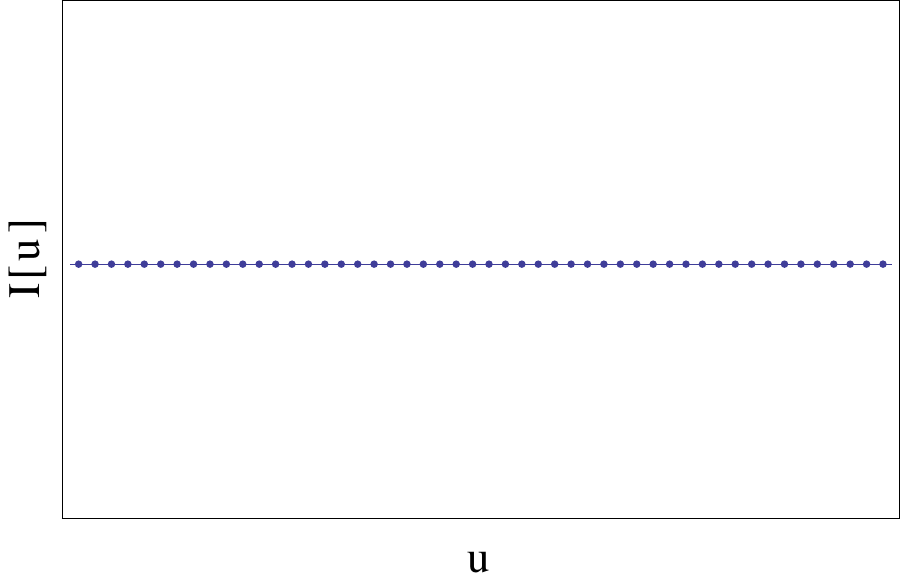}
\caption{Test of the approximations used in the analytical calculation for scattering with the modes at $\pi$ and $\pi(1\pm2 M)$. The integral $I(u)=\int_p\int_q \frac{\delta(\xi_{\sigma\vec p}) \delta(\xi_{\sigma'\vec q+\vec p+Q})  }{N_\sigma N_{\sigma'}} |V(p)|^2|V(q+p+Q)|^2  \im{\chi^R(u,q_\|+Q)}$ is numerically evaluated (dots) and the scaling compared to the analytical expression $I(u)\propto u^{1-2\gamma}$ (solid) for $\gamma=3/4$ and $\gamma=1/2$.}
\label{fig:numSE}
\end{figure}

In the main text, we have considered the contribution to the self-energy due to scattering with staggered, commensurate ($S_+S_-$) and incommensurate ($S_zS_z$) spin modes. The corresponding spin susceptibilities are given by
\begin{align}
 \im{\chi_{\s\s'}^{R}(\nu,q_\|)}&=\chi_{\s\s'}^{0} \sin(\gamma_{\s\s'} \pi)\text{sgn}(\nu)\frac{\Theta(\nu^2-v_s^2q_\|^2)}{|-\nu^2+v_s^2q_\|^2|^{\gamma_{\s\s'}}}
 \end{align}
with $\gamma_{\s,\bar\s}=1-1/(4K)$ ($\bar\s=-\s$) and $\gamma_{\s,\s}=1-K$. To obtain the corresponding frequency dependence of the self-energy, we use the following analytical approximations
\begin{align}\label{eq:SEpiInt}
\Gamma_{\sigma\sigma'}^c(\omega)&=\frac{J^2 \chi_{+-}^{c,0}\sin\gamma\pi}{2} \int du (\sgn{u}-\sgn{u+\omega})\int_p\int_q  \sgn{u} \frac{\Theta(u^2-v_s^2 q_\|^2)}{|u^2-v_s^2 q_\|^2|^\gamma}  |V(p)|^2|V(q+p+Q)|^2  \frac{\delta(\xi_{\sigma \vec p}) \delta(\xi_{\bar\sigma \vec p+\vec q +Q})}{N_\sigma N_{\bar \sigma}}  \notag \\
&\approx \frac{J^2 \chi_{+-}^{c,0}\sin\gamma\pi}{2} \int du (1-\sgn{u}\sgn{u+\omega}) \frac{1}{|u|^{2\gamma}} \int_p\int_q \Theta(u^2-v_s^2 q_\|^2) |V(p)|^2|V(q+p+Q)|^2  \frac{\delta(\xi_{\sigma \vec p}) \delta(\xi_{\bar\sigma \vec p+\vec q +Q})}{N_\sigma N_{\bar \sigma}}  \notag \\
&\approx \frac{J^2 \chi_{+-}^{c,0}\sin\gamma\pi}{2} \int du (1-\sgn{u}\sgn{u+\omega}) \frac{2|u|}{v_s|u|^{2\gamma}} \int_p\int_q  |V(p)|^2|V(q+p+Q)|^2  \frac{\delta(\xi_{\sigma \vec p}) \delta(\xi_{\bar\sigma \vec p+\vec q +Q})}{N_\sigma N_{\bar \sigma}} \delta(q_\|)\notag \\
&\approx \frac{ \chi_{+-}^{c,0}\sin\gamma\pi}{2} \int du (1-\sgn{u}\sgn{u+\omega}) \frac{2|u|}{v_s|u|^{2\gamma}} \widehat{ V^2}\notag \\
&=\frac{ \chi_{+-}^{c,0}\sin\gamma\pi}{v_s} \frac{1}{1-\gamma}\widehat{ V^2} |\omega|^{2-2\gamma} \notag \\
&\propto \left\{ \begin{matrix} |\omega|^{\frac{1}{2K}} & \s'=\s \\ |\omega|^{2K} & \s'=-\s \end{matrix} \right.
 \end{align}
 with $\widehat{ V^2}=J^2\int_p\int_q  |V(p)|^2|V(q+p+Q)|^2  \frac{\delta(\xi_{\sigma \vec p}) \delta(\xi_{\bar\sigma \vec p+\vec q +Q})}{N_\sigma N_{\bar \sigma}}$.
Here, we have discarded terms of higher order in frequency than the leading term $\omega^{2-2\gamma}$ and approximated the prefactor by the average of the squared coupling due to simplicity. To check our approximations, we have calculated, the momentum integral in the first line numerically. The result is shown in Fig.~\ref{fig:numSE}. Indeed, we obtain the correct scaling with our analytical analysis.

Let us also note that formally, we have to distinguish
\begin{align}
\chi_{\s\s}^{s}(i\omega,q_\|)&=\chi_{\s\s}^{s,0}\frac{1}{(\omega^2+v_s^2q_{||}^2)^{1-K}} \quad \text{if } K<1\\
\chi_{\s\s}^{s}(i\omega,q_\|)&=\chi_{\s\s}^{s,0}\ln\frac{(v_s \Lambda)^2}{\omega^2+v_s^2q_{||}^2} \quad \text{if } K=1,
\end{align}
In the case  $K\approx1$ of incommensurate $S_zS_z$ modes; the imaginary part of the spin susceptibility is then
 \be
\im{\chi_{z}^{ic,R}(\nu,q)}=\chi_{z}^{ic,0} \pi\text{sgn}(\nu)\Theta(\nu^2-v_s^2q^2)
\ee
Hence we find for the incommensurate modes around $\pi(1\pm 2M)$ 
\begin{align}
\Gamma_{\sigma\sigma}^{ic}(\omega)&\approx\frac{J^2 \chi_{z}^{ic,0}\pi}{2} \int (1-\sgn{u}\sgn{u+\omega})\frac{2|u|}{v_s} \int_p\int_q  |V(p)|^2|V(q+p+Q)|^2  \frac{\delta(\xi_{\sigma \vec p}) \delta(\xi_{\sigma \vec p+\vec q +Q})}{N_\sigma^2}  \notag \\
&=\frac{J^2 \chi_{z}^{ic,0}\pi}{v_s}\omega^2 \int_p\int_q  |V(p)|^2|V(q+p+Q)|^2  \frac{\delta(\xi_{\sigma \vec p}) \delta(\xi_{\sigma \vec p+\vec q +Q})}{N_\sigma^2}
 \end{align}
 which coincides with the limit $K\rightarrow 1$ of the expression for general $K$.

\subsection{Relaxation rate from uniform spin correlations} \label{app:uniform}
To perform the  integrals in the general expression for the electron self-energy, we have to take the different spin susceptibilities into account. Here we present the calculations for the subleading modes, which have been omitted in the main text. We will use the same approximations as for the staggered modes, which we checked numerically above.

For the commensurate $S_zS_z$ correlations, we have 
\be
\im{\chi_z^{c,R}(\nu,q)}=\chi_{z}^{c,0}\pi \text{sgn}(\nu)q_z^2\delta(-\nu^2+v_s^2q^2)
\ee
For zero temperature, this leads to
\begin{align}
\Gamma_{\sigma\sigma}^c(\omega)&=\frac{J^2 \chi_z^{c,0}\pi}{2} \int du (\sgn{u}-\sgn{u+\omega})\int_p\int_q q_\|^2 \sgn{u} \delta (u^2-v_s^2 q_\|^2) \left( \frac{p_\|^2+p_{\perp,1}^2}{p^2} \right)^3 \left( \frac{(q_\|+p_\|)^2+(q_{\perp,1}+p_{\perp,1})^2}{(q+p)^2} \right)^3 \notag \\ & \hspace{6.5cm}\times\frac{\delta(\xi_{\sigma \vec p}) \delta(\xi_{\sigma \vec p+\vec q })}{N_\sigma^2}  \notag \\
&=\frac{J^2 \chi_z^{c,0}\pi}{2} \int du (1-\sgn{u}\sgn{u+\omega})\int_p\int_q \frac{|q_\||}{2v_s^2} \left(\delta(q_\|-\frac{u}{v_s})+\delta(q_\|+\frac{u}{v_s})\right) \left( \frac{p_\|^2+p_{\perp,1}^2}{p^2} \right)^3 \notag \\
& \hspace{6.5cm} \times\left( \frac{(q_\|+p_\|)^2+(q_{\perp,1}+p_{\perp,1})^2}{(q+p)^2} \right)^3 \frac{\delta(\xi_{\sigma \vec p}) \delta(\xi_{\sigma \vec p+\vec q })}{N_\sigma^2}  
\notag \\
&=\frac{J^2 \chi_z^{c,0}\pi}{2} \int_0^{|\omega|} du 2 \frac{|u|}{2v_s^3}\int_p\int \frac{d^2 q_\perp}{(2\pi)^3}  \!\left( \frac{p_\|^2+p_{\perp,1}^2}{p^2} \right)^3 \!\! \frac{\delta(\xi_{\sigma \vec p}) }{N_\sigma^2}\! \left[|V(q+p)|^2 \delta(\xi_{\sigma \vec p+\vec q })|_{q_\|=\frac{u}{v_s}}+|V(q+p)|^2\delta(\xi_{\sigma \vec p+\vec q })|_{q_\|=-\frac{u}{v_s}} \right]
\notag \\
&\approx \frac{J^2 \chi_z^{c,0}\pi}{2v_s^3} \omega^2 \int_p \int_q |V(p)|^2|V(q+p)|^2 \frac{\delta(\xi_{\sigma \vec p}) \delta(\xi_{\sigma \vec p+\vec q })}{N_\sigma^2}
\end{align} 

Similarly, to lowest order in $\omega$, the FS-averaged relaxation rate due to incommensurate $S_+S_-$ modes around $\pm2\pi M$ with 
\be
\im{\chi_{+-}^{ic,R}(\nu,q)}=\chi_{+-}^{ic,0} \sin(\beta\pi)\text{sgn}(\nu)\frac{\nu^2+v_s^2q_\|^2}{|-\nu^2+v_s^2q^2|^{\beta}}\Theta(\nu^2-v_s^2q^2) \quad  \quad \quad \beta=2-K-\frac{1}{4K}
\ee
is
\begin{align}
\Gamma_{\sigma\bar\sigma}^{ic}(\omega)&=\frac{J^2 \chi_{+-}^{ic,0}\sin\beta\pi}{2} \int du (\sgn{u}-\sgn{u+\omega})\int_p\int_q  \sgn{u} \frac{u^2+v_s^2q_\|^2}{|u^2-v_s^2 q_\|^2|^\beta} \Theta(u^2-v_s^2 q_\|^2) |V(p)|^2|V(q+p+Q)|^2 \notag \\ 
&\hspace{7cm} \times \frac{\delta(\xi_{\sigma \vec p}) \delta(\xi_{\bar\sigma \vec p+\vec q +Q})}{N_\sigma N_{\bar \sigma}}  \notag \\
&\approx \frac{J^2 \chi_{+-}^{ic,0}\sin\beta\pi}{2} \int du (1-\sgn{u}\sgn{u+\omega})\frac{2|u|}{v_s} \frac{u^2}{|u|^{2\beta}} \int_p\int_q   |V(p)|^2|V(q+p+Q)|^2  \frac{\delta(\xi_{\sigma \vec p}) \delta(\xi_{\bar\sigma \vec p+\vec q +Q})}{N_\sigma N_{\bar \sigma}}  \notag \\
&=\frac{J^2 \chi_{+-}^{ic,0}\sin\beta\pi}{v_s} \frac{1}{2-\beta} |\omega|^{4-2\beta} \int_p\int_q  |V(p)|^2|V(q+p+Q)|^2  \frac{\delta(\xi_{\sigma \vec p}) \delta(\xi_{\bar\sigma \vec p+\vec q +Q})}{N_\sigma N_{\bar \sigma}}  \notag \\
&\propto |\omega|^{2K+\frac{1}{2K}}
 \end{align}
 In the Luttinger liquid regime, $K$ varies between $1/4$ and $1$ and for an extended magnetic field regime it is close to $1/2$. Thus, the leading contribution to the full relaxation rate $\Gamma_\sigma(\omega)=\sum_{i,\sigma'}N_{\sigma'}\int du (n_B(u)+n_F(u+\omega))\Gamma_{\sigma\sigma'}^i$ comes from $\Gamma_{\sigma\sigma}^{ic}$ and $\Gamma_{\sigma\bar\sigma}^{c}$, i.e. the staggered  $S_zS_z$  and $S_+S_-$ modes around $Q\sim\pi(1\pm2\pi M)$ and $Q\sim\pi$, respectively. This is why we focus on them in the main text.

\subsection{Adding the polarization bubble}
\label{app:withbubble}
When adding the polarization bubble to the bare spin propagator, i.e. $\chi^{-1}(\nu,q)=\chi^{-1}_{bare}(\nu,q)+\Pi(\nu,q)$ with $\Pi(\nu,q)\approx -\Pi_r - i\tau_{\vec q_\perp} \nu$, the imaginary part of the spin propagator becomes 
\begin{align}
\im \chi^{i,R}_{\sigma\sigma'}(\nu,q)=&\chi_{\sigma\sigma'}^{i,0}\,\Theta(\nu^2-v_s^2q^2) \frac{\tilde\tau_{\vec q_\perp} \nu + \text{sgn}(\nu)\sin(\gamma_i \pi)|-\nu^2+v_s^2q^2|^{\gamma_i}}{\Big(- \tilde\Pi_r+\cos(\gamma_i\pi)|-\nu^2+v_s^2q^q|^{\gamma_i} \Big)^2+\Big( \tilde\tau_{\vec q_\perp}\nu+\text{sgn}(\nu)\sin(\gamma_i\pi)|-\nu^2+v_s^2q^q|^{\gamma_i} \Big)^2} \notag\\
+&\chi_{\sigma\sigma'}^{i,0}\,\Theta(v_s^2q^2-\nu^2) \frac{\tilde\tau_{\vec q_\perp}\nu}{\tilde\tau_{\vec q_\perp}^2\nu^2+\Big(-\tilde\Pi_r+|-\nu^2+v_s^2q^2|^{\gamma_i}\Big)^2}
\end{align}
with $\gamma_i=\gamma_{\s\s},\gamma_{\s\bar\s}$ and $\tilde \tau=\chi_0\tau$, $\tilde \Pi_r=\chi_0\Pi_r$.
The contribution from small internal frequencies in the lower row induces behavior linear in frequency. For sufficiently small $\tau\nu$, this contribution looks like a Delta distribution and we obtain
\begin{align}
\Gamma_{\sigma\sigma'}&=\frac{J^2\chi_{\sigma\sigma'}^{i,0}\pi}{2}\int du (1-\sgn{u}\sgn{u+\omega})\int_p\int_q\Theta(v_s^2q_\|^2-u^2)\delta(-\tilde \Pi_r+|u^2-v_s^2q_\|^2|^{\gamma_i})|V(p)|^2|V(p+q+Q)|^2\notag \\
&\hspace{6.5cm}\times \frac{\delta(\xi_{\sigma \vec p}) \delta(\xi_{\sigma' \vec p+\vec q +Q})}{N_\sigma N_{ \sigma'}} \notag \\
&=\frac{J^2\chi_{\sigma\sigma'}^{i,0}\pi}{2}\int du (1-\sgn{u}\sgn{u+\omega})\int_p\int_q\Theta(v_s^2q_\|^2-u^2) \frac{\delta(q_\|+\frac{1}{v_s}\sqrt{\tilde \Pi_r^{1/\gamma_i}+u^2})+\delta(q_\|-\frac{1}{v_s}\sqrt{\tilde \Pi_r^{1/\gamma_i}+u^2})}{2v_s\gamma_i \tilde \Pi_r^{1-1/\gamma_i}\sqrt{\tilde \Pi_r^{1/\gamma_i}+u^2}}  \notag \\
&\hspace{6.5cm}\times |V(p)|^2|V(p+q+Q)|^2\frac{\delta(\xi_{\sigma \vec p}) \delta(\xi_{\sigma' \vec p+\vec q +Q})}{N_\sigma N_{ \sigma'}} \notag \\
&\approx \frac{J^2\chi_{\sigma\sigma'}^{i,0}\pi}{2}\int du 2\frac{1-\sgn{u}\sgn{u+\omega}}{2v_s\gamma_i\tilde \Pi_r^{1-1/(2\gamma_i)}}\int_p\int_q |V(p)|^2|V(p+q+Q)|^2\frac{\delta(\xi_{\sigma \vec p}) \delta(\xi_{\sigma' \vec p+\vec q +Q})}{N_\sigma N_{ \sigma'}} \notag \\
&= \frac{J^2\chi_{\sigma\sigma'}^{i,0}\pi}{v_s\gamma_i\tilde \Pi_r^{1-1/(2\gamma_i)}}|\omega|\int_p\int_q |V(p)|^2|V(p+q+Q)|^2\frac{\delta(\xi_{\sigma \vec p}) \delta(\xi_{\sigma' \vec p+\vec q +Q})}{N_\sigma N_{ \sigma'}}
\end{align}
We see that the result scales linearly in frequency.

\subsection{Conductivity}
\label{app:cond}

\subsubsection{General Form}

The conductivity is given by the retarded current-current correlation function (we are interested in the diagonal compontents)
\begin{align}
\sigma_a(\omega,\vec q)&=\frac{i}{\omega}\Pi^R_{j_a}(\omega,\vec q),\\
\Pi_{j_a}(i\omega,\vec q)&=-\frac{1}{V}\int_0^\beta d\tau e^{i\omega \tau}\la T_\tau j_a(\tau,\vec q) j_a(0,\vec q)\ra
\end{align}
with volume $V$, time ordering $T_\tau$ and $a$-component of the current operator $\vec j$. The dc conductivity is obtained by taking the limits $q\rightarrow 0$, and then $\omega\rightarrow0$ of the expression with the retarded current correlation function (order of limits is important). In our case, we have to distinguish the current for spin up and down $\vec j=\vec j_\uparrow+\vec j_\downarrow$, so that we consider $\sigma=\sigma_\uparrow+\sigma_\downarrow=\frac{i}{\omega}\left[\Pi^R_{j,\uparrow}(\omega,\vec q)+\Pi^R_{j,\downarrow}(\omega,\vec q)\right]$. Let us define the current vertex $\vec \Gamma_a^\sigma= p_a \gamma_\sigma$. Then the static current-current correlation function becomes
\begin{align}
\pi_a^\sigma(i\omega)&=\lim_{q\rightarrow 0} \Pi_{j_a,\sigma}(i\omega,\vec q) \notag\\
&=\frac{2e^2}{m^2}\int \frac{d^3p}{(2\pi)^3}T\sum_{ip} G_\sigma(ip,\vec p)G_\sigma(ip+i\omega,\vec p)  p_a \Gamma_a^\sigma (\vec p,ip,ip+i\omega) \notag \\
&=\frac{2e^2}{m^2}\int \frac{d^3p}{(2\pi)^3} p_a^2T\sum_{ip} G_\sigma(ip,\vec p)G_\sigma(ip+i\omega,\vec p) \gamma_\sigma (\vec p,ip,ip+i\omega)
\end{align}
In our case, we want to distinguish directions parallel and perpendicular to the chains, i.e. we consider
\begin{align}
\pi_\|^\sigma(i\omega)&=\frac{2e^2}{m^2}\int \frac{d^3p}{(2\pi)^3} p_\|^2T\sum_{ip} G_\sigma(ip,\vec p)G_\sigma(ip+i\omega,\vec p) \gamma_\sigma (\vec p,ip,ip+i\omega)\\
\pi_\perp^\sigma(i\omega)&=\frac{e^2}{m^2}\int \frac{d^3p}{(2\pi)^3} \vec p_\perp^2T\sum_{ip} G_\sigma(ip,\vec p)G_\sigma(ip+i\omega,\vec p) \gamma_\sigma (\vec p,ip,ip+i\omega),
\end{align}
where the factor of $1/2$ is added because both perpendicular directions yield the same result so that $\pi^\sigma_\perp=\pi^\sigma_x=\pi^\sigma_y$.

After evaluating the Matsubara summation, the dc conductivity becomes (Ref. \cite{Mahan} Ch. 8.4.2)
\begin{align}
\sigma_{\|/\perp}^\sigma=&\frac{1}{1+\delta_{\perp,\|/\perp}}\frac{e^2}{m^2}\int \frac{d^3p}{(2\pi)^2}p_{\|/\perp}^2\int \frac{d\epsilon}{2\pi}\left(-\frac{dn_F }{d\epsilon}\right) \notag \\
&\times\left[ G_\sigma(\vec p,\epsilon+i\delta) G_\sigma(\vec p,\epsilon-i\delta) \gamma_\sigma(\vec p,\epsilon-i\delta,\epsilon+i\delta) +\text{Re}\Big(G^2_\sigma(\vec p,\epsilon+i\delta) \gamma_\sigma(\vec p,\epsilon+i\delta,\epsilon + i\delta)\Big) \right]
\end{align}
The second term in the brackets will vanish by doing the integration over $\xi=p^2/(2m)-\mu$, which is why we will not consider it in the following.
Furthermore, we use that due to the Green's functions, the dominant contribution comes from momenta close to the FS and that we are only interested in scaling with frequency or temperature, i.e. we can replace $|G_\sigma|^2\approx \delta(\xi_{\sigma\vec p})/\Gamma(\epsilon)$. This leads to
\begin{align}
\sigma^\sigma_i&\approx \frac{2e^2}{c_i m^2}\int \frac{d\epsilon}{2\pi} \left(-\frac{dn_F }{d\epsilon}\right) \int_p p_i^2 \frac{\delta(\xi_{\sigma\vec p})}{\Gamma_\s(\epsilon)} \gamma_\sigma(\vec p, \epsilon) \notag \\
&=\frac{2e^2}{c_i m^2}\int \frac{d\epsilon}{2\pi} \left(-\frac{dn_F }{d\epsilon}\right)  \frac{1}{\Gamma_\s(\epsilon)} p_{F\sigma}^2 \int_p \frac{p_i^2}{p^2} \delta(\xi_{\sigma\vec p}) \gamma_\sigma(\vec p, \epsilon) \notag \\
&=\frac{6e^2n_\sigma}{c_i m}\int \frac{d\epsilon}{2\pi} \left(-\frac{dn_F }{d\epsilon}\right)  \frac{1}{\Gamma_\s(\epsilon)} \frac{1}{N_\sigma} \int_p \frac{p_i^2}{p^2} \delta(\xi_{\sigma\vec p}) \gamma_\sigma(\vec p, \epsilon)\notag \\
&=:\frac{6e^2n_\sigma}{c_i m}\int \frac{d\epsilon}{2\pi} \left(-\frac{dn_F }{d\epsilon}\right)  \frac{\gamma^i_\sigma(\epsilon)}{\Gamma_\s(\epsilon)} 
\end{align}
with $c_i=1+\delta_{i\perp}$, $\gamma_\sigma(\vec p, \epsilon)=\gamma_\sigma(\vec p, \epsilon-i\delta,\epsilon+i\delta)$, density of states $N_\sigma=p_{F\sigma}m/(2\pi^2)$ and number density $n_\sigma=p_{F\sigma}^3/(6\pi^2)$. We have defined the FS-averaged, scalar current vertex $\gamma^i_\sigma(\epsilon)=\frac{1}{N_\sigma} \int_p \frac{p_i^2}{p^2} \delta(\xi_{\sigma\vec p}) \gamma_\sigma(\vec p, \epsilon)$. 
We observe that
\be
\gamma_\sigma(\epsilon)=\frac{1}{N_\sigma} \int_p \delta(\xi_{\sigma\vec p}) \gamma_\sigma(\vec p, \epsilon)=\frac{1}{N_\sigma} \int_p \frac{p_\|^2+\vec p_\perp^2}{p^2} \delta(\xi_{\sigma\vec p}) \gamma_\sigma(\vec p, \epsilon)=\gamma^\|_\sigma(\epsilon)+ \gamma^\perp_\sigma(\epsilon)
\ee

\subsubsection{Vertex function}
To obtain the scalar vertex functions $\gamma_\sigma^{\perp/\|}$, we sum the ladder diagrams for the current vertex. This leads to the Dyson equation for the vertex of the form
\begin{align}
\vec p\gamma_\sigma(\vec p,ip,ip+i\omega)=&\vec p+ J^2 \sum_{\sigma'} T\sum_{iq}\int_q |V(p)|^2 |V(q)|^2 \chi_{\sigma\sigma'}(iq-ip,\vec q-\vec p) G(iq,\vec q)G(iq+i\omega,\vec q) \vec q \gamma_{\sigma'}(\vec q,iq,iq+i\omega).
\end{align}
After performing the Matsubara summation and again approximating $|G_\sigma|^2\approx \frac{\delta(\xi_{\vec p\sigma})}{\Gamma_\s(\epsilon)}$, the equation for $\gamma_\sigma(\vec p, \epsilon-i\delta,\epsilon+i\delta)$ becomes
\begin{align}
\vec p\gamma_\sigma(\vec p,\epsilon)=&\vec p+ J^2 \sum_{\sigma'} \int dx \int_q |V(p)|^2 |V(q)|^2 \im\chi^R_{\sigma\sigma'}(x,\vec q-\vec p) \left[n_B(x) +n_F(\epsilon+x) \right] \frac{\delta(\xi_{\vec q\sigma'})}{\Gamma_{\s'}(\epsilon+x)} \vec q \gamma_{\sigma'}(\vec q,\epsilon+x).
\end{align}
Thus the averages over the Fermi surface satisfy
\begin{align}
\gamma_\sigma^i(\epsilon)&=\int_p \frac{\delta(\xi_{\vec p\sigma})}{N_\sigma}\frac{p_i^2}{p^2}\gamma_\sigma(\vec p,\epsilon)\notag \\
&=m_i+ J^2 \sum_{\sigma'}N_{\sigma'}\int dx\int_p \int_q |V(p)|^2 |V(q)|^2 \im\chi^R_{\sigma\sigma'}(x,\vec q-\vec p) \left[n_B(x) +n_F(\epsilon+x) \right] \frac{\delta(\xi_{\vec p\sigma})\delta(\xi_{\vec q\sigma'})}{N_\sigma N_{\sigma'}} \frac{\vec p_i \vec q_i}{p^2} \frac{\gamma_{\sigma'}(\vec q,\epsilon+x)}{\Gamma_{\s'}(\epsilon+x)}.
\end{align}
with 
\be
m_i=\int_p \frac{\delta(\xi_{\vec p\sigma})}{N_\sigma}\frac{p_i^2}{p^2}
\ee
and 
\begin{align}\label{eq:gamma}
\gamma_\sigma(\epsilon)&=
\int_p \frac{\delta(\xi_{\vec p\sigma})}{N_\sigma}\gamma_\sigma(\vec p,\epsilon) \notag \\
&=1+ J^2 \sum_{\sigma'}N_{\sigma'}\int dx\int_p \int_q |V(p)|^2 |V(q)|^2 \im\chi^R_{\sigma\sigma'}(x,\vec q-\vec p) \left[n_B(x) +n_F(\epsilon+x) \right] \frac{\delta(\xi_{\vec p\sigma})\delta(\xi_{\vec q\sigma'})}{N_\sigma N_{\sigma'}} \frac{\vec p \vec q}{p^2} \frac{\gamma_{\sigma'}(\vec q,\epsilon+x)}{\Gamma_{\s'}(\epsilon+x)} \notag \\
&=1+ J^2 \sum_{\sigma'}N_{\sigma'}\int dx\int_p \int_q |V(p)|^2 |V(q)|^2 \im\chi^R_{\sigma\sigma'}(x,\vec q-\vec p) \left[n_B(x) +n_F(\epsilon+x) \right] \frac{\delta(\xi_{\vec p\sigma})\delta(\xi_{\vec q\sigma'})}{N_\sigma N_{\sigma'}} \frac{\vec p \vec q}{p^2} \notag \\
&\hspace{3cm}\times \frac{1}{\Gamma_{\s'}(\epsilon+x)} \int dx\int_q \frac{\delta(\xi_{\vec q\sigma'})}{N_{\sigma'}}\gamma_{\sigma'}(\vec q,\epsilon+x) \notag \\
&=1+   \sum_{\sigma'}N_{\sigma'} \int  du [n_B(u)+n_F(u+\epsilon)]I_{\sigma\sigma'}(u) \frac{\gamma_{\sigma'}(\epsilon+u)}{\Gamma_{\sigma'}(\epsilon+u)}.
\end{align}
In the last line we have defined $I_{\sigma\sigma'}(u)=J^2\int_p \int_q |V(p)|^2 |V(q)|^2 \im\chi^R_{\sigma\sigma'}(u,\vec q-\vec p) \frac{\delta(\xi_{\vec p\sigma})\delta(\xi_{\vec q\sigma'})}{N_\sigma N_{\sigma'}} \frac{\vec p \vec q}{p^2}$. 
The integral $I_{\sigma\sigma'}$ can be rewritten as a sum of relaxation rate and a transport rate
\begin{align}
I_{\sigma\sigma'}=J^2\int_p \int_q |V(\vec p)|^2 |V(\vec q+ \vec p +Q_{\sigma\sigma'})|^2 \im\chi^R_{\sigma\sigma'}(u,\vec q+Q_{\sigma\sigma'}) \frac{\delta(\xi_{\vec p\sigma})\delta(\xi_{\vec q+\vec p + Q\sigma'})}{N_\sigma N_{\sigma'}} \frac{\vec p (\vec q+\vec p +Q_{\sigma\sigma'})}{p^2}
\end{align}
with $Q_{\sigma\sigma}=\pi(1\pm2\pi M)$ and $Q_{\sigma\bar\sigma}=\pi$. The product $(p-(q+p+Q))^2= p_{F\sigma}^2+p_{F\sigma'}^2-2p(q+p+Q)=(Q+q)^2=Q^2+2Qq+q^2$, so that
\begin{align}
I_{\sigma\sigma'}&=J^2\int_p \int_q |V(\vec p)|^2 |V(\vec q+ \vec p +Q_{\sigma\sigma'})|^2 \im\chi^R_{\sigma\sigma'}(u,\vec q+Q_{\sigma\sigma'}) \frac{\delta(\xi_{\vec p\sigma})\delta(\xi_{\vec q+\vec p + Q\sigma'})}{N_\sigma N_{\sigma'}} \notag \\
&\hspace{5cm}\times\left(\frac{p_{F\sigma}^2+p_{F\sigma'}^2-Q_{\sigma\sigma'}^2}{2p_{F\sigma}^2} -\frac{q(2Q_{\sigma\sigma'}+q)}{2p_{F\sigma}^2}\right) \notag \\
&=f_{\sigma\sigma'} \Gamma_{\sigma\sigma'}- \Gamma_{\sigma\sigma'}^t
\end{align}
with $f_{\sigma\sigma'}=\frac{p_{F\sigma}^2+p_{F\sigma'}^2-Q_{\sigma\sigma'}^2}{2p_{F\sigma}^2}\approx 1-Q^2/(2p_{F}^2)+\mathcal{O}(h/E_F)$ and $\Gamma_{\sigma\sigma'}^t=J^2\int_p \int_q |V(\vec p)|^2 |V(\vec q+ \vec p +Q)|^2 \im\chi^R_{\sigma\sigma'}(u,\vec q+Q) \frac{\delta(\xi_{\vec p\sigma})\delta(\xi_{\vec q+\vec p + Q\sigma'})}{N_\sigma N_{\sigma'}} \frac{q(2Q+q)}{2p_{F\sigma}^2}$. To solve the equation for the current vertex $\gamma_\sigma(\epsilon)$, we approximate $\frac{\gamma_{\sigma'}(\epsilon+x)}{\Gamma_{\sigma'}(\epsilon+x)}\rightarrow \frac{\gamma_{\sigma'}(\epsilon)}{\Gamma_{\sigma'}(\epsilon)}$, which has been shown to lead to the correct low-energy scaling \cite{belitz2010}.  Due to the additional factor in the transport relaxation rate $\Gamma_{\s\s'}^t$, the leading frequency dependence comes from $I_{\sigma\sigma'}\approx f_{\sigma\sigma'}\Gamma_{\sigma\sigma'}$.  
With this, the scalar current vertex is given by
\begin{align}\label{eq:fullcurrvert}
\gamma_\sigma&=\Gamma\frac{\Gamma+f_{\s\bar\s}\Gamma_{+-}-f_{\bar\s\bar\s}\Gamma_z}
{\Gamma^2-f_{\s\bar\s}f_{\bar\s\s}\Gamma_{+-}^2 -(f_{\s\s}+f_{\bar\s\bar\s}) \Gamma\Gamma_z +f_{\s\s}f_{\bar\s\bar\s} \Gamma_z^2} \notag \\
&\approx \frac{1}{1-f}+\mathcal{O}(h/E_F),
\end{align}
where we have neglected terms of order of the magnetic field over the Fermi energy $h/E_F\ll1$ and $f=1-Q^2/(2p_{F}^2)$. We cited this result in the main text.
Note that this is only the leading contribution for a finite $Q$: if $Q\approx 0$ and $f\rightarrow 1$, this contribution vanishes and the leading behavior is of the form $\gamma_\s=\Gamma_\s/\Gamma^t_\s$ with  $ \Gamma^t_{\s}=\sum_{\sigma'}N_{\sigma'} \int  du [n_B(u)+n_F(u+\epsilon)]\Gamma^t_{\sigma\sigma'}(u)$.
\end{appendix}

\end{widetext}

\end{document}